\documentclass[conference]{IEEEtran}
\IEEEoverridecommandlockouts
\usepackage{cite}
\usepackage{amsmath,amssymb,amsfonts}
\usepackage{algorithmic}
\usepackage{graphicx}
\usepackage{textcomp}
\usepackage{xcolor}
\usepackage{url}
\usepackage{subfigure}
\usepackage{float}
\usepackage[top=0.55in, bottom=0.55in, left=0.635in, right=0.635in]{geometry}
\def\BibTeX{{\rm B\kern-.05em{\sc i\kern-.025em b}\kern-.08em
    T\kern-.1667em\lower.7ex\hbox{E}\kern-.125emX}}
\begin{document}

\title{Age-Oriented Opportunistic Relaying in Cooperative Status Update Systems with Stochastic Arrivals\\{\thanks{This work is done when the first author Bohai Li is a visiting student at the Chinese University of Hong Kong.}
	}
}

\author{\IEEEauthorblockN{Bohai Li\textsuperscript{1,2}, He Chen\textsuperscript{2}, Yong Zhou\textsuperscript{3}, and Yonghui Li\textsuperscript{1}}%
	\IEEEauthorblockA{\textsuperscript{1}School of Electrical and Information Engineering, The University of Sydney, Sydney, Australia}
	\IEEEauthorblockA{\textsuperscript{2}Department of Information Engineering, The Chinese University of Hong Kong, Hong Kong SAR, China}
	\IEEEauthorblockA{\textsuperscript{3}School of Information Science and Technology, ShanghaiTech University, Shanghai, China\\
	\textsuperscript{1}\{bohai.li, yonghui.li\}@sydney.edu.au,
	\textsuperscript{2}he.chen@ie.cuhk.edu.hk, \textsuperscript{3}zhouyong@shanghaitech.edu.cn}
}

\maketitle

\begin{abstract} This paper considers a cooperative status update system with a source aiming to send randomly generated status updates to a designated destination as timely as possible with the help of a relay. We adopt a recently proposed concept, Age of Information (AoI), to characterize the timeliness of the status updates. We propose an age-oriented opportunistic relaying (AoR) protocol to reduce the AoI of the considered system. Specifically, the relay opportunistically replaces the source to retransmit the successfully received status updates that have not been correctly delivered to the destination, but the retransmission of the relay can be preempted by the arrival of a new status update at the source. By carefully analyzing the evolution of AoI, we derive a closed-form expression of the average AoI for the proposed AoR protocol. We further minimize the average AoI by optimizing the generation probability of the status updates at the source. Simulation results validate our theoretical analysis and demonstrate that the average AoI performance of the proposed AoR protocol is superior to that of the non-cooperative system.



\end{abstract}

\section{Introduction}
With the rapid development of the Internet of Things (IoT) applications, timely status updates have become increasingly critical in many areas, such as telesurgery and self-driving cars \cite{b1}. However, the conventional performance metrics, e.g., throughput and delay, cannot properly characterize the timeliness of the status updates \cite{b2}. For example, throughput can be maximized by generating and transmitting the status updates as frequent as possible. However, excessive update rates may lead to network congestion, which makes the updates suffer from long transmission delay. Such long delay can be reduced by lowing the update rate. However, reducing the rate of updates overmuch can also make the monitor receive undesired outdated status updates. Motivated by this, age of information (AoI), defined as the time elapsed since the generation of the latest received status update, has been recently introduced to quantify the information freshness from the perspective of the receiver that monitors a remote process \cite{b3}. Unlike conventional metrics, AoI is related to both the transmission delay and the update generation rate \cite{b4}. As a result, AoI is a more comprehensive evaluation criterion for information freshness. 


Since the AoI concept was first proposed to characterize the information freshness in a vehicular status update system \cite{b5}, extensive work focusing on the analysis and optimization of AoI has appeared. Most of the existing work is concerned with the AoI of single-hop wireless networks \cite{b6,b7, b8, b9, b10, b11, b12, b13}. On the other hand, the AoI performance in multi-hop networks has also been studied in \cite{b14, b15, b16, b17}. However, all the above work overlooked the direct link between the source and the destination. Therefore, the updates from the source can only be transmitted to the destination via the relay. As far as we know, there is no existing work that designs and optimizes the average AoI of a cooperative status update system with the existence of a direct link. Such a design is indeed non-trivial. This is because delivering the status updates via the direct link will have shorter transmission time at the cost of a higher error probability, while the delivery of status updates through the two-hop relay link could be more reliable at the cost of longer transmission time.

Motivated by this gap, in this paper we consider a three-node cooperative status update system, where a source timely reports randomly generated status updates to its destination with the help of a relay. We concentrate on designing and analyzing an age-oriented opportunistic relaying (AoR) protocol to reduce the AoI of the considered system. By carefully observing the AoI evolution process, we first define some necessary time intervals to mathematically express the average AoI of the proposed protocol. By representing these time-interval definitions in terms of key system parameters, including the generation probability of the status updates and the transmission success probabilities of three links, we then attain a closed-form expression of the average AoI for the proposed AoR protocol. Given reasonable values of the transmission success probabilities, we further minimize the average AoI by optimizing the status generation probability at the source. Simulation results are then provided to validate the theoretical analysis. Furthermore, the simulation results demonstrate that the average AoI of the proposed AoR protocol is smaller than that of the non-cooperative system, especially when the source-destination link has bad channel conditions.

%

\section{System Model and Protocol Description}
We consider a three-node cooperative status update system, in which the source (S) wants to send randomly generated status updates to a designated destination (D) as timely as possible with the help of a relay (R). There is a direct link between S and D. The transmission of the status updates can either go through the direct link or the S-R-D link. To quantify the timeliness of the status updates, we adopt a recently proposed metric, termed AoI, which is defined as the time elapsed since the generation of the last successfully received status update. It is assumed that the considered system is time slotted and the transmission of each status update by S/R takes exactly one time slot, whose length is normalized to one without loss of generality. All the channels suffer from block fading, i.e., the channels responses remain unchanged within one time slot but vary independently from one time slot to another.

\subsection{Protocol Description}
We now propose an AoR protocol for the considered system. S transmits a newly generated status update to D. If the status update is successfully received by D, then an acknowledgement (ACK) packet is sent back to S to discard the status update at the end of the current time slot. Otherwise, either S or R retransmits the status update to reduce the average AoI. Specifically, if R successfully receives the status update, then R feeds back an ACK to inform S to discard the status update, and transmits the status update to D by itself in the following time slot(s). Note that the channel gain of the R-D link is generally better than that of the S-D link. Retransmitting the same status update packets by R instead of S can reduce the AoI. If R fails to receive the status update, then S keeps retransmitting until either D or R succeeds in receiving the said status update. To ensure that fresh status updates can be transmitted timely, we enforce S to preempt the retransmissions from R in case of a new status update arrival. Hence, if R has a status update to retransmit, it first senses the state of S at the beginning of one time slot. If S with a new status update transmits, then R discards its current status packet and attempts to decode the new one sent by S. Otherwise, R keeps retransmitting the update until the successful reception at D or the preemption by S. For simplicity, we assume that both the time for sensing and transmitting an ACK packet are negligible. 

It is worth mentioning that in the conventional cooperation protocols that are designed mainly from a physical layer perspective, normally allocate dedicated channel resources for R to facilitate the cooperation \cite{b18}. However, if the transmission success probability of the S-D link is large, it is apparent that transmitting more status updates through the S-D link reduces the AoI. Therefore, conventional cooperation protocols become sub-optimal from the perspective of minimizing the AoI. We note that a novel cooperation protocol, where R utilizes the silence periods of S terminals to enable cooperation, was proposed in \cite{b19} to improve the cooperative system performance. However, the design in \cite{b19} focused on the throughout maximization rather than AoI minimization. As far as we know, the AoR protocol is the first effort towards the design and analysis for the considered system from the perspective of AoI.


\begin{figure}[tbp]
	\centerline{\includegraphics[height=5.8cm]{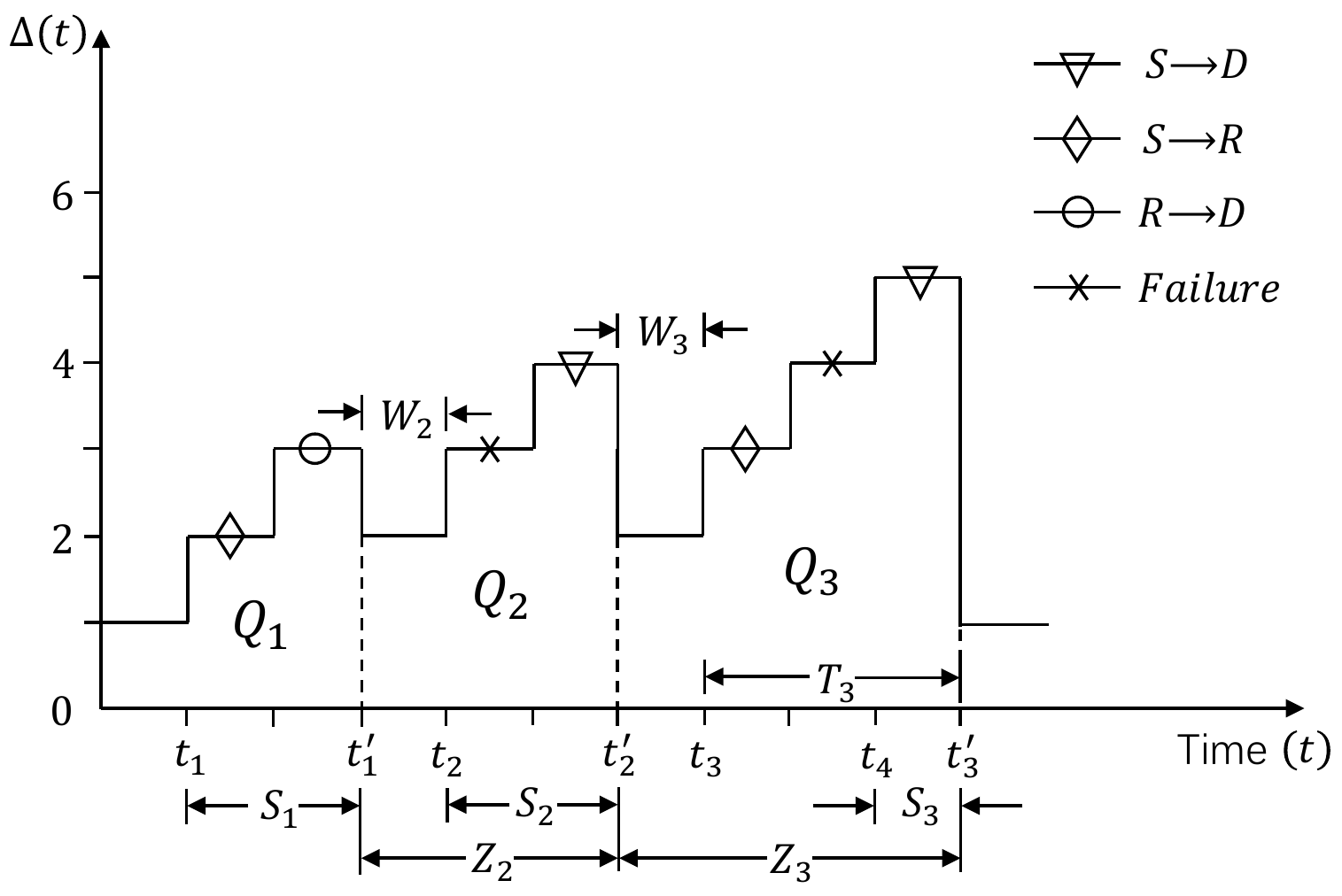}}
	\vspace{-0.7 em}
	\caption{Sample staircase path of the AoI evolution. We use $S\!\rightarrow\! D$, $S\!\rightarrow\! R$, $R\!\rightarrow\! D$ and Failure to denote the successful transmission through the S-D link, the successful transmission through the S-R link, the successful transmission through the R-D link and the failed transmission, respectively.}
	\vspace{-1.2 em}
	\label{fig:two-hoop-relay-model-1}
\end{figure}

\subsection{AoI Definition}
We follow \cite{b6}, \cite{b8} and adopt a Bernoulli process to model the stochastic arrivals at S. Particularly, a new status update is generated with probability $p$ at the beginning of each time slot. We denote $t_i$ as the generation time of the $i$th update at S. Taking into account channel fading, we define by $P_{1}$, $P_{2}$, and $P_{3}$ to denote the transmission success probabilities through the S-D link, S-R link and R-D link, respectively. Due to the transmission error and the aforementioned preemption, the status updates generated at S may not always be successfully received by D. Hence, we further denote $t'_{k}$ as the arrival time of the $k$th status update successfully received by D. We denote $U(t)$ as the generation time of the most recently received status update at D until time slot $t$. The AoI at time slot $t$ can then be defined as 
\vspace{-0.1cm}
\begin{equation}
\Delta(t) = t - U(t).
\vspace{-0.1cm}	
\end{equation}

For the ease of understanding the AoI evolution, we illustrate an example staircase path for 10 consecutive time slots with an initial value of one in Fig. 1. Note that we also define some time intervals in Fig. 1 to facilitate the calculation of the AoI, which will be explained one by one in the following. We denote by $N_{k}$ the index of the most recently generated update at S before $t'_{k}$, i.e., 
\begin{equation}
N_{k}=\max\{i\!\mid\!t_{i} < t'_{k}\},
\end{equation}
and denote by $N'_{k}$ the index of the first generated update at S since the last successfully received update at D, i.e., 
\begin{equation}
N'_{k}=\min\{i\!\mid\!t_{i} > t'_{k-1}\}.
\end{equation}
We further define $S_{k}$ as the service time of the $k$th successfully received update at D, given by
\begin{equation}
S_{k} = t'_{k} - t_{N_{k}}.
\end{equation}
In addition, $W_{k}$ is defined as the waiting time starting from the arrival of the $(k-1)$th successfully received update at D until the generation of the first status update at S after $t'_{k-1}$, which is given by 
\begin{equation}
W_{k}=t_{N'_{k}}-t'_{k-1}.
\end{equation}
We define $T_{k}$ as the time duration between the generation time of the first status update after $t'_{k-1}$ and the arrival time of the $k$th successfully received update at D, which is given by
\begin{equation}
T_{k}=t'_{k}-t_{N'_{k}},	
\end{equation}
and $Z_{k}$ is defined as interdeparture time between two consecutive successfully received status updates at D, given by
\begin{equation}
Z_{k}=t'_{k}-t'_{k-1}.
\end{equation}

In the following section, we shall analyze the average AoI of the proposed AoR protocol for the considered cooperative status update system.



\section{Analysis and Optimization of Average AoI}

In this section, we first express the average AoI by the defined time intervals given in Section II. By representing these time-interval definitions in terms of key system parameters, (i.e., $p$, $P_{1}$, $P_{2}$ and $P_{3}$), we then manage to attain a closed-form expression of the average AoI for the proposed AoR protocol. Given reasonable value sets of $P_{1}$, $P_{2}$ and $P_{3}$, we further minimize the average AoI by optimizing the status generation probability $p$.

\subsection{Analysis of Average AoI}
\vspace{-0.04cm}
We define $X_{t}$ as the number of successfully received updates at D till time $t$ and it can be expressed as 
\vspace{-0.075cm}
\begin{equation}
	X_{t} = \max\{k\!\mid\!t'_{k} <t\}.
	\vspace{-0.075cm}	
\end{equation}

\noindent As depicted in Fig. 1, the average AoI can be expressed by the area under the AoI curve, i.e., $Q_{k}$, and mathematically we have 
\vspace{-0.02cm}
\begin{equation}
	\bar{\Delta}=\lim\limits_{t \to \infty }{\frac{X_{t}}{t}\frac{1}{X_{t}}\sum_{k=1}^{X_{t}}Q_{k}}=\frac{\mathbb{E}[Q_{k}]}{\mathbb{E}[Z_{k}]}.
\end{equation} 
\vspace{-0.31cm}

\noindent For further simplification, we represent $Q_{k}$ in terms of the time intervals defined in Section II and it can be expressed as
\vspace{-0.13cm}
\begin{equation}
	\begin{aligned}
		Q_{k} &=S_{k-1}+(S_{k-1}+1)+\cdots+(S_{k-1}+Z_{k}-1)\\
		&= S_{k-1}Z_{k}+\frac{Z_{k}^2-Z_{k}}{2}.
	\end{aligned}
	\vspace{-0.09cm}
\end{equation}

\noindent By substituting (10) into (9), we can express the average AoI as
\vspace{-0.05cm} 	
\begin{equation}
\bar{\Delta}=\frac{\mathbb{E}[S_{k-1}Z_{k}]}{\mathbb{E}[Z_{k}]}+\frac{\mathbb{E}\Big[Z_{k}^2\Big]}{2\mathbb{E}[Z_{k}]}-\frac{1}{2}.
\vspace{-0.05cm} 
\end{equation}
In order to further simplify (11), we have the following lemma.

\textit{\underline{Lemma} 1:} The service time of the $(k-1)$th successfully received update at D is independent of the interdeparture time between the $(k-1)$th and the $k$th successfully received updates at D such that:
\vspace{-0.1cm}
\begin{equation}
\mathbb{E}[S_{k-1}Z_{k}]=\mathbb{E}[S_{k-1}]\mathbb{E}[Z_{k}].
\end{equation}

\vspace{-0.13cm}
The lemma can be proved by following a similar process as that for \cite[Lemma 1]{b6}

By applying Lemma 1, the expression of the average AoI can be simplified as
\vspace{-0.22cm}
\begin{equation}
\bar{\Delta}=\mathbb{E}[S_{k-1}]+\frac{\mathbb{E}\Big[Z_{k}^2\Big]}{2\mathbb{E}[Z_{k}]}-\frac{1}{2}.
\end{equation}

\vspace{-0.1cm}
To obtain the average AoI, we now derive the terms $\mathbb{E}[S_{k-1}]$, $\mathbb{E}[Z_{k}]$ and $\mathbb{E}[Z_{k}^2]$ one by one in the following. We start with the expectation of the service time, i.e., $\mathbb{E}[S_{k-1}]$. In the considered system, there are two types of status updates that can be successfully received by D without being preempted. One are these packets that are successfully received by D through the direct link, and the other are these packets that are successfully delivered through the two-hop S-R-D link. Therefore, $\mathbb{E}[S_{k-1}]$ can be expressed as a weighted sum of the average service time of each type. For the first type of successful updates, the probability that the update is successfully received after $l$ times of transmission by S can be expressed as  
\vspace{-0.12cm}	
\begin{equation}
P_{l} = (1-p)^{l-1}(1-P_{2})^{l-1}(1-P_{1})^{l-1}P_{1}.
\vspace{-0.15cm}	
\end{equation}
The corresponding expectation of the service time can be calculated by
\begin{equation}
\mathbb{E}_{l}=\frac{\sum_{l=1}^{\infty}P_{l}\cdot l}{\sum_{l=1}^{\infty}P_{l}}.
\end{equation}
Similarly, for the second type of successful status update, the expectation of its service time can be expressed by
\begin{equation}
\mathbb{E}_{mn}=\frac{\sum_{n=1}^{\infty}\sum_{m=1}^{\infty}P_{mn}\cdot(m+n)}{\sum_{n=1}^{\infty}\sum_{m=1}^{\infty}P_{mn}},
\end{equation} 

\noindent where $m$ and $n$ denote the number of transmission times via the S-R link and the R-D link, respectively. $P_{mn}$ is the probability that the update is successfully received by D after transmitting $(m+n)$ times, which can be expressed as
\begin{equation}
\resizebox{.98\hsize}{!}{$P_{mn} \!=\! (1-p)^{m-1}(1-P_{2})^{m-1}(1-P_{1})^{m}P_{2}(1-p)^{n}(1-P_{3})^{n-1}P_{3}$}.
\end{equation}
Since the above two types of updates make up all the updates that can be successfully received by D, the expectation of the service time of the considered system can be evaluated as 
\begin{equation}
\resizebox{.8\hsize}{!}{$\begin{aligned}
	\mathbb{E}[S_{k-1}]=&\ \mathbb{E}_{l}\cdot\frac{\sum_{l=1}^{\infty}P_{l}}{\sum_{l=1}^{\infty}P_{l}+\sum_{n=1}^{\infty}\!\sum_{m=1}^{\infty}\!P_{mn}}\\
	&\,+ \mathbb{E}_{mn}\cdot\frac{\sum_{n=1}^{\infty}\!\sum_{m=1}^{\infty}\!P_{mn}}{\sum_{l=1}^{\infty}P_{l}+\sum_{n=1}^{\infty}\!\sum_{m=1}^{\infty}\!P_{mn}},
	\end{aligned}
	$}
\end{equation}

\noindent which follows according to the law of total probability.

By applying the finite sum equations given in \cite[Eqs. (0.112) and (0.113)]{b20}, we can further simplify (18) to
\begin{equation}
\mathbb{E}[S_{k-1}]=\frac{1}{1-\beta}+\frac{1}{1-\alpha}\cdot\frac{\gamma}{P_{1}(1-\alpha)+\gamma},
\end{equation}
where $\alpha=(1-p)(1-P_{3})$, $\beta=(1-p)(1-P_{1})(1-P_{2})$ and $\gamma=P_{2}P_{3}(1-p)(1-P_{1})$. 

Based on (5), (6) and (7), we have $Z_{k}=W_{k}+T_{k}$. Therefore, the expectation of the interdeparture time can be written as $\mathbb{E}[Z_{k}]=\mathbb{E}[W_{k}]+\mathbb{E}[T_{k}]$. Recall that S can preempt the transmission of R when there is a new arrival. This indicates that the waiting time $W_{k}$ only depends on the generation probability $p$. Since the status updates are generated according to a Bernoulli process, $W_{k}$ follows a geometric distribution with parameter $p$ and its expectation can be readily given by $\mathbb{E}[W_{k}]=(1-p)/p$. We then move to the calculation of term $\mathbb{E}[T_{k}]$.

Note that $T_{k}$ behaves differently for the following four possible cases: 1) The update is successfully received by D through the S-D link without being preempted; 2) The update is successfully received by D through the S-R-D link without being preempted; 3) The update is preempted by a new update before being successfully received by either R or D, and the new update may be preempted by multiple new updates; 4) The update is preempted by a new update generated at S after being successfully received by R, and the new update may be preempted by multiple new updates at S. We note that the number of preemptions can approach infinity in the third and fourth cases, which generally makes the expectation of $T_{k}$ difficult to derive mathematically. To tackle this issue, we resort to the recursive method applied in \cite{b6},\cite{b21} and evaluate the expectation of $T_{k}$ as (20) at the top of the next page.
\begin{figure*}
	\begin{equation}
	\resizebox{.79\hsize}{!}{$\begin{aligned}
		\mathbb{E}[T_{k}]=&\
		\sum_{l=1}^{\infty}P_{l}\cdot l + \sum_{n=1}^{\infty}\sum_{m=1}^{\infty}P_{mn}\cdot (m+n)
		+ \sum_{l=1}^{\infty}(1-P_{1})^{l}(1-P_{2})^{l}(1-p)^{l-1}p\cdot \big(l+\mathbb{E}\big[T'_{k}\big]\big)\\ &\,+\sum_{m=1}^{\infty}\sum_{n=0}^{\infty}\Big[(1-P_{1})^{m}(1-P_{2})^{m-1}P_{2}(1-p)^{m-1}
		\times(1-p)^{n}(1-P_{3})^{n}p\cdot \big(m+n+\mathbb{E}\big[T'_{k}\big]\big)\!\Big]\\
		\end{aligned}$}
	\end{equation}
	\noindent\rule[0.05\baselineskip]{\textwidth}{0.5pt}
	\vspace{-1cm}
\end{figure*}

Note that the four terms on the right hand side of (20) correspond to the above four cases, respectively, and $T'_{k}$ is the time duration between the preemption time (i.e., the generation time of the second status update after $t'_{k-1}$) and the arrival time of the $k$th successfully received update at D. As $T_{k}$ is defined as the time duration between the generation time of the first status update after $t'_{k-1}$ and the arrival time of the $k$th successfully received update at D, following the idea of recursion, we have $\mathbb{E}[T_{k}]=\mathbb{E}\big[T'_{k}\big]$. After some manipulations by applying \cite[Eqs. (0.112) and (0.113)]{b20}, we have
\vspace{-0.2cm}
\begin{equation}
\mathbb{E}[T_{k}]=\frac{(1-\alpha)+\beta\cdot P'_{2}}{P_{1}(1-\alpha)+\gamma},
\vspace{-0.1cm}
\end{equation}
where $P'_{2} = P_{2}/(1-P_{2})$. 
By combining $\mathbb{E}[W_{k}]=(1-p)/p$ and (21), we now attain a closed-form expression for the term $\mathbb{E}[Z_{k}]$, given by
\vspace{-0.1cm}
\begin{equation}
\begin{aligned}
\mathbb{E}[Z_{k}]=\mathbb{E}[W_{k}]+\mathbb{E}[T_{k}]=\frac{(1-\alpha)(1-\beta)}{p\big[P_{1}(1-\alpha)+\gamma\big]}.
\end{aligned}
\vspace{-0.1cm}
\end{equation}

Finally, we work on the expectation of $Z_{k}^{2}$, which can be expressed as 
\vspace{-0.05cm}
\begin{equation}
\begin{aligned}
\mathbb{E}\Big[Z_{k}^{2}\Big]=\mathbb{E}\Big[(W_{k}+T_{k})^2\Big]=\mathbb{E}\Big[W_{k}^{2}\Big]+2\mathbb{E}[W_{k} T_{k}]+\mathbb{E}\Big[T_{k}^{2}\Big].
\end{aligned}
\end{equation}
It is readily to find that $W_{k}$ and $T_{k}$ are independent. Thus, we have $\mathbb{E}[W_{k}T_{k}]=\mathbb{E}[W_{k}]\mathbb{E}[T_{k}]$. As $W_{k}$ follows a geometric distribution with parameter $p$, we have $\mathbb{E}[W_{k}^2]=(p^2-3p+2)/p^2$. In order to evaluate $\mathbb{E}[Z_{k}^2]$, the only remaining task is to calculate $\mathbb{E}[T_{k}^2]$.

Similar to (20), the expectation of $T_{k}^2$ can be evaluated as (24) at the top of the next page.
\begin{figure*}
	\begin{equation}
	\resizebox{.9\hsize}{!}{$\begin{aligned}
		\mathbb{E}\Big[T_{k}^{2}\Big]\!=\!\!&\
		\sum_{l=1}^{\infty}P_{l}\!\cdot\! l^2 \!+\! \sum_{n=1}^{\infty}\!\sum_{m=1}^{\infty}\!P_{mn}\!\cdot\! (m^2\!+\!2mn\!+\!n^2)
		\!+\!\sum_{l=1}^{\infty}\!\bigg[\!(1\!-\!P_{1})^{l}(1\!-\!P_{2})^{l}(1\!-\!p)^{l-1}p
		\!\cdot\!\Big(l^2\!+\!2l\mathbb{E}\big[T_{k}\big]\!+\!\mathbb{E}\Big[T_{k}^{2}\Big]\Big)\!\bigg]\\
		&\,\!+\! \sum_{m=1}^{\infty}\!\sum_{n=0}^{\infty}\!\bigg[\!(1\!-\!P_{1})^{m}(1\!-\!P_{2})^{m-1}P_{2}(1\!-\!p)^{m-1}
		(1\!-\!p)^{n}(1\!-\!P_{3})^{n}p
		\!\cdot\! \Big(\!m^2\!+\!n^2\!+\!\mathbb{E}\Big[T_{k}^{2}\Big]\!
		+\!2mn\!+\!2m\mathbb{E}[T_{k}]\!+\!2n\mathbb{E}[T_{k}]\!\Big)\!\bigg]
		\end{aligned}$}
	\end{equation}
	\noindent\rule[0\baselineskip]{\textwidth}{0.5pt}
	\vspace{-0.95cm}
\end{figure*}


Similar to the process of obtaining (21) from (20), we can rewrite $\mathbb{E}[T_{k}^2]$ as
\vspace{-0.15cm}
	\begin{equation}
\resizebox{.95\hsize}{!}{$\begin{aligned}
	\mathbb{E}\Big[T_{k}^2\Big]\!= &\,\frac{1}{(1-\alpha)(1-\beta)-(1-\alpha)\beta\cdot p'-\beta\cdot p'\cdot P'_{2}}\\
	&\times\!\frac{1}{(1-\alpha)(1-\beta)}\\
	&\times\!\bigg\{(1-\alpha)^2(1+\beta)+(3-\alpha-\beta-\alpha\beta)\beta\cdot P'_{2}\\
	&\qquad\!\!+2\mathbb{E}[T_{K}]\!\cdot\!\Big[(1\!-\!\alpha)^2\beta\!\cdot \!p'\!+\!(1\!-\!\alpha\beta)\beta\!\cdot\! p'\!\cdot\! P'_{2} \Big]\bigg\},
	\end{aligned}$}
\end{equation}
where $p'=p/(1-p)$.

By substituting the terms derived in (19), (22), (23) and (25) into (13), we can obtain the exact closed-form expression of the average AoI for the proposed AoR protocol, given by
\vspace{-0.1cm}
\begin{equation}
\bar{\Delta}\!=\!\frac{\big[1\!-\!(1\!-\!p)(1\!-\!P_{3})\big]\!\cdot\! \big[1\!-\!(1\!-\!p)(1\!-\!P_{1})(1\!-\!P_{2})\big]}{p\!\cdot\!\big[pP_{1}\!+\!(1\!-\!p)P_{3}\!-\!(1\!-\!p)(1\!-\!P_{1})(1\!-\!P_{2})P_{3}\big]}.
\end{equation}
\subsection{Optimization of Average AoI}
In this subsection, given reasonable value sets of $P_{1}$, $P_{2}$ and $P_{3}$, we will minimize the average AoI by optimizing the generation probability $p$ at S.

By assuming that $0<P_{1}<P_{2}<1$ and $0<P_{1}<P_{3}<1$, the optimal generation probability $p$ that minimizes the average AoI is presented in the following theorem.

\textit{\underline{Theorem} 1:} The optimal generation probability that minimizes the average AoI is given by: 
\begin{equation}
	p=\left\{
	\begin{aligned}
	&\frac{-\!\psi\!+\!\sqrt{\psi^2\!-\!4\chi \omega}}{2\chi},\\
	&\quad \mathrm{if} \ 
	0<P_{1}<\frac{P_{2}\!+\!P_{2}P_{3}\!-\!\sqrt{(P_{2}\!-\!P_{2}P_{3})^2\!+\!4P_{2}P_{3}}}{2(P_{2}\!-\!1)},\\
	&1, \\
	&\quad \mathrm{if} \ 
	\frac{P_{2}\!+\!P_{2}P_{3}\!-\!\sqrt{(P_{2}\!-\!P_{2}P_{3})^2\!+\!4P_{2}P_{3}}}{2(P_{2}\!-\!1)}<P_{1}<1,
	\end{aligned} \right.
	\end{equation} 
\noindent where $\chi\!=\!P_{2}P_{3}(1\!-\!P_{2}P_{3})\!-\!P_{1}^2(1\!-\!P_{2})(1\!-\!P_{3})^2\!-\!P_{1}P_{2}P_{3}^2(1\!-\!P_{2})(2\!-\!P_{1})\!-\!P_{1}P_{2}(1\!-\!P_{3})$, $\psi\!=\!-2P_{3}(P_{1}\!+\!P_{2}\!-\!P_{1}P_{2})(P_{1}\!-\!P_{1}P_{3}\!-\!P_{2}P_{3}\!+\!P_{1}P_{2}P_{3})$ and $\psi^2\!-\!4\chi \omega\!=\!4P_{2}P_{3}^2(P_{3}\!-\!P_{1})(1\!-\!P_{1})(P_{1}\!+\!P_{2}\!-\!P_{1}P_{2})^2$.

Proof: See Appendix A.

\textit{\underline{Remark} 1:} 
For the second case in Theorem 1, the average AoI is minimized by setting $p=1$ and it is given by 
\begin{equation}
\bar\Delta_{min}=\frac{1}{P_{1}}.
\end{equation}
Note that in the considered system, the minimum possible time required for a successful status update through the S-D link is one time slot, while that through the S-R-D link is at least two time slots. Therefore, when the transmission success probability of the S-D link (i.e., $P_{1}$) is relatively large, having more status updates successfully delivered through the S-D link is beneficial to reduce the average AoI. If the generation probability of the updates is 1, S always has a new status update to transmit (i.e., the generate-at-will model \cite{b22}), which means that the updates successfully received by R in the previous time slot is always preempted. As a result, all the successfully received status updates by D can only come from the S-D link, which will minimize the average AoI.


On the other hand, for the first case in Theorem 1, the transmission success probability of the S-D link is relatively small compared to the second case. Hence, the updates successfully delivered through the S-R-D link also have a significant contribution on minimizing the average AoI. As mentioned above, if the generation probability is 1, the updates can only be successfully delivered to D through the S-D link. Obviously, this is not optimal for minimizing the average AoI in this case, which means that $p=1$ should not be the optimal status generation probability to minimize the average AoI. 

\section{Simulation Results and Discussions}

In this section, we present simulation results to validate our analytical results, and demonstrate the superiority of the considered cooperative status update system over its non-cooperative counterpart in terms of achieving lower average AoI. For the non-cooperative status update system, we plot its average AoI curves according to the result given in \cite[Eqs. (35)]{b6}. Each curve presented in the section is averaged over $10^7$ time slots.

We first study the average AoI of both cooperative and non-cooperative systems against the the transmission success probability of the S-D link (i.e., $P_{1}$), as shown in Fig. 2, for two different generation probabilities of the status updates. In both cases, we assume that the transmission success probabilities of the S-R link and the R-D link are $P_{2}\!\!=\!\!P_{3}\!\!=\!\!0.8$. It can be observed that in both cases, the performance of the considered cooperative status update system is obviously superior to that of the non-cooperative system, especially when $P_{1}$ is small. Specifically, when $p\!=\!0.5$ (or $p\!=\!0.9$), the considered cooperative system has a significant performance improvements over the non-cooperative system for the case $P_{1}\!<\!0.5$ (or $P_{1}\!<\!0.3$), respectively.
This is because when the S-D link has bad channel conditions, the generated updates at S can be transmitted to D via R in our system while those in the non-cooperative system can only be retransmitted via the direct link. However, we can also observe that the AoI performance gain decreases as $P_{1}$ increases, and approaches 1 when $P_1$ is close to 0.8 (i.e., equal to $P_{2}$ and $P_{3}$). The rationality behind this is that when $P_{1}$ is large, most of the successfully received updates at D are transmitted via the S-D link. Therefore, R is not much helpful in reducing the average AoI. We also note that when $P_{1}$ is less than 0.27, a lower generation probability, i.e., $p\!=\!0.5$, results in a better AoI performance. On the contrary, the AoI performance of a higher generation probability, i.e., $p\!=\!0.9$, is always better when $P_{1}$ is greater than 0.27. Motivated by this interesting observation, we explore the relationship between the generation probability and the average AoI in the following.

We now investigate the average AoI against the generation probability $p$ by fixing the transmission success probabilities of the S-R link and the R-D link as $P_{2}\!=\!P_{3}\!=\!0.8$. As depicted in Fig. 3, we also consider two cases with the transmission success probability of the S-D link being 0.25 and 0.5, respectively. In the first case of $P_{1}=0.25$, it is apparent that the average AoI first decreases and then increases as the generation probability increases. This is because if the S-D link has bad channel conditions, the updates successfully delivered to D by R have a significant contribution on minimizing the average AoI. However, as S can preempt the transmission of R whenever there is a new arrival, if the updates are generated too frequently, the updates at R will be always preempted by S and cannot be transmitted to D to effectively reduce the AoI. Therefore, too high generation probabilities in turn increase the average AoI. We can also find that the average AoI when $p=0.5$ is smaller than that when $p=0.9$, which confirms our mentioned observation when $P_{1}$ is less than 0.27 in Fig. 2. In the second case of $P_{1}=0.5$, the average AoI always decreases as the generation probability increases. The monotone decreasing property demonstrates our mentioned phenomenon when $P_{1}$ is greater than 0.27 in Fig. 2. It is also worth mentioning that the optimal generation probabilities to minimize the average AoI in the case of $P_{1}\!=\!0.25$ and $P_{1}\!=\!0.5$ are $p=0.6770$ and $p=1$, respectively, which is consistent with the results calculated by using the formula in Theorem 1.

\begin{figure}[tbp]
	\centering{\includegraphics[height=5.8cm]{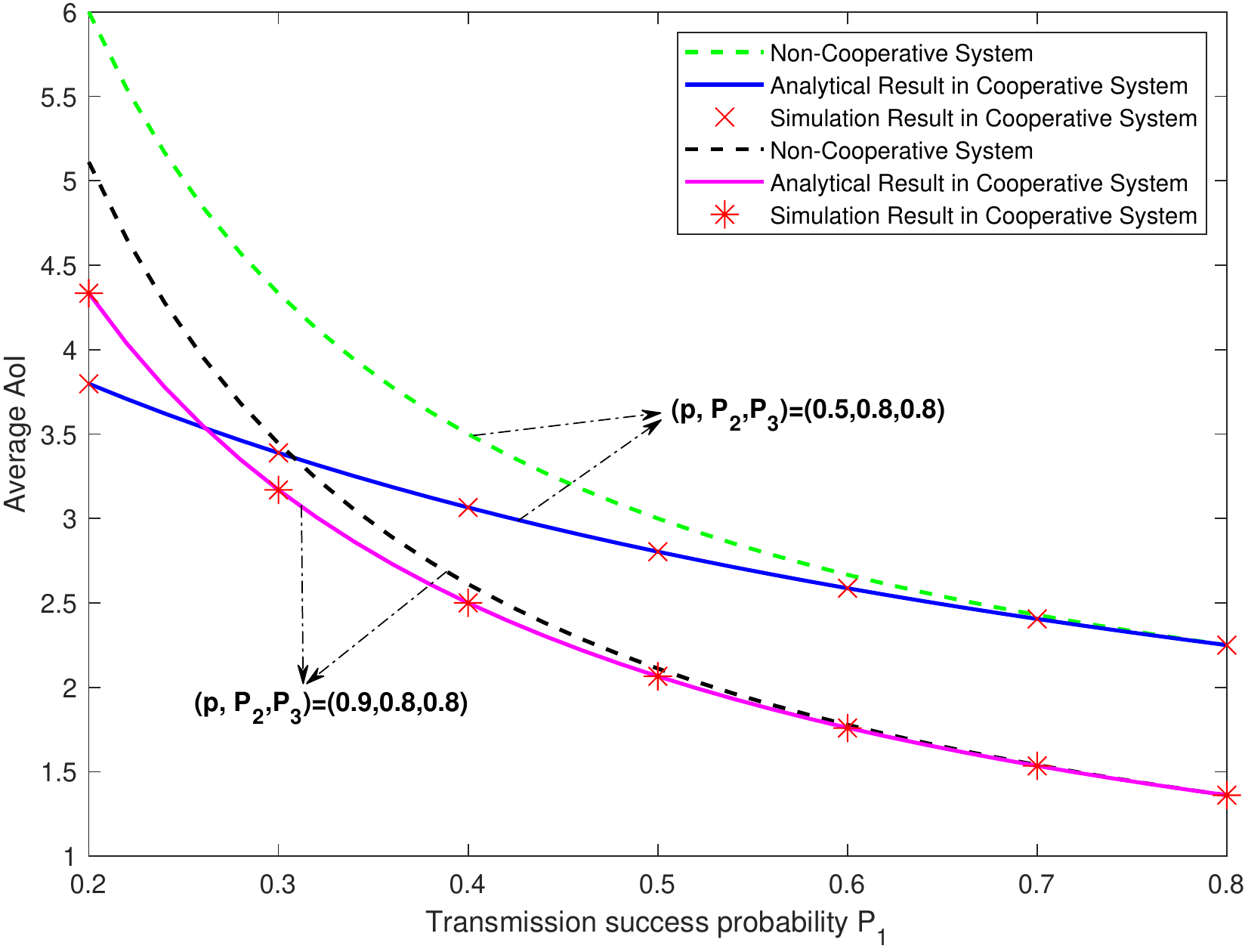}}
	\vspace{-0.3cm}
	\caption{Average AoI of the considered cooperative system and the non-cooperative system versus the transmission success probability $P_{1}$ for different generation probabilities.}
	\vspace{-0.3 cm}
	\label{fig:two-hoop-relay-model-1}
\end{figure}
\begin{figure}[tbp]
	\centering{\includegraphics[height=5.8cm]{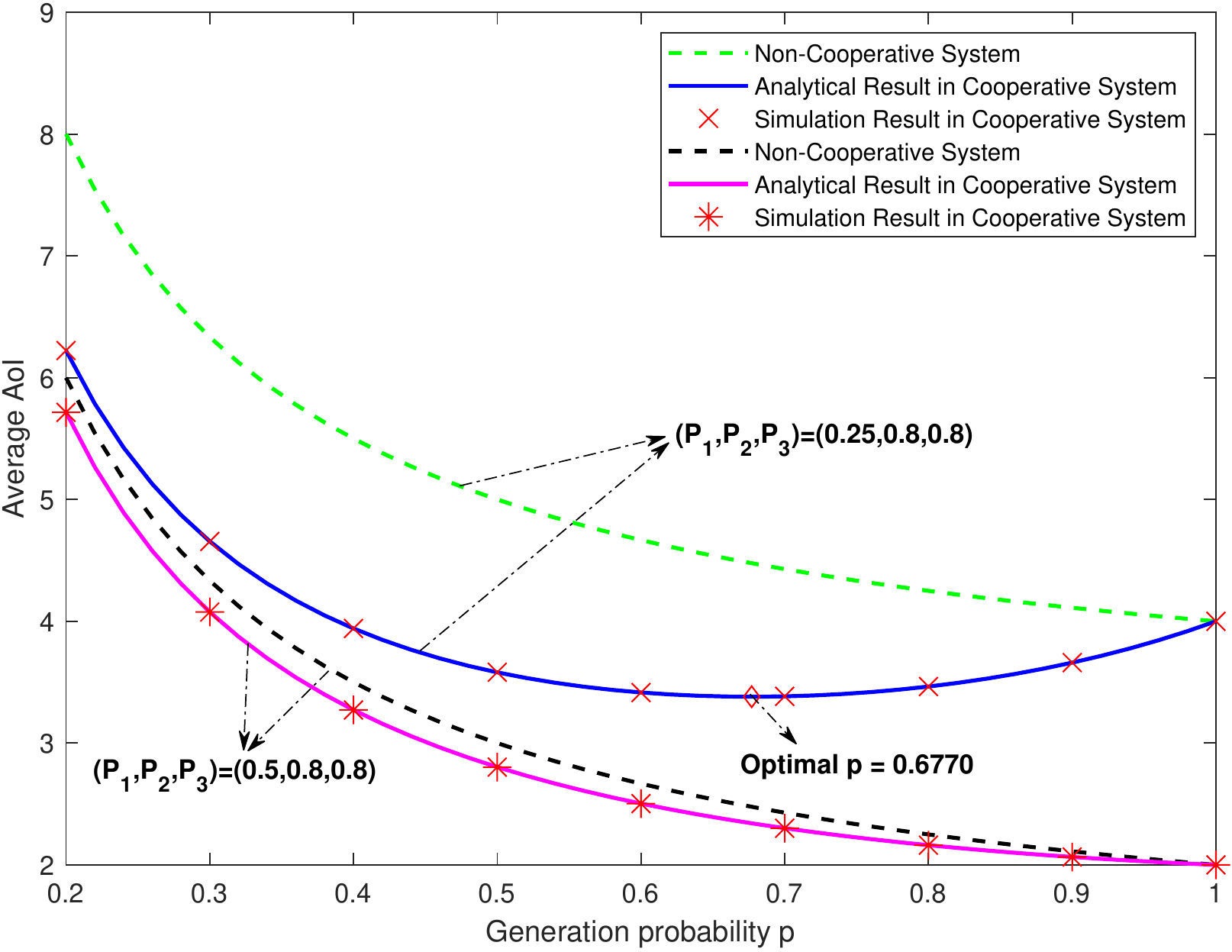}}
	\vspace{-0.2 cm}
	\caption{Average AoI of the considered cooperative system and the non-coopertive system versus the generation probability $p$ for different transmission success probabilities, i.e., $(P_{1}, P_{2}, P_{3})$.}
	\vspace{-0.6 cm}
	\label{fig:two-hoop-relay-model-1}
\end{figure}

\vspace{-0.2 cm}
\section{Conclusions}
\vspace{-0.15cm}
In this paper, we proposed an age-oriented opportunistic relaying (AoR) protocol to reduce the age of information (AoI) of a cooperative status update system, where the source timely reports randomly generated status updates to the destination with the help of a relay. By analyzing the evolution of AoI, we derived a closed-form expression of the average AoI of the proposed AoR protocol as a function of the generation probability of the status updates and the transmission success probability of each link. Given reasonable assumptions about the transmission success probabilities, we further figured out the optimal generation probability of the status updates that minimizes the average AoI. Simulation results validated our theoretical analysis and demonstrated the superiority of the proposed AoR protocol over the non-cooperative system in terms of minimizing the average AoI, especially when the source-destination link is not as good as the other two links.

\section*{Appendix}
%
%

\textit{A. Proof of Theorem 1}

Recall that $\bar{\Delta}=\frac{[1-(1-p)(1-P_{3})]\cdot [1-(1-p)(1-P_{1})(1-P_{2})]}{p[pP_{1}+(1-p)P_{3}-(1-p)(1-P_{1})(1-P_{2})P_{3}]}$. To proceed, we derive the first-order derivative of $\bar\Delta$ with respect to (w.r.t) $p$. After some algebra manipulations, we have
\begin{equation}
\frac{\partial \bar{\Delta}}{\partial p}=\frac{\chi \cdot p^2+ \psi \cdot p+\omega}{p^2\!\cdot\!\big[pP_{1}\!+\!(1\!-\!p)P_{3}\!-\!(1\!-\!p)(1\!-\!P_{1})(1\!-\!P_{2}P_{3})\big]^2},
\end{equation}
where
\begin{subequations}
	\begin{equation}
	\resizebox{.99\hsize}{!}{$\begin{aligned}
		\chi\!=
		\!-&(P_{1}^2P_{2}^2P_{3}^2\!-\!2P_{1}^2P_{2}P_{3}^2\!+\!2P_{1}^2P_{2}P_{3}\!-\!P_{1}^2P_{2}\!+\!P_{1}^2P_{3}^2\!
		-\!2P_{1}^2P_{3}\!\\	
		&\,+\!P_{1}^2\!
		-\!2P_{1}P_{2}^2P_{3}^2\!+\!2P_{1}P_{2}P_{3}^2\!-\!P_{1}P_{2}P_{3}\!+\!P_{1}P_{2}\!+\!P_{2}^2P_{3}^2\!-\!P_{2}P_{3}),
		\end{aligned}$}
	\end{equation}
	\begin{equation}
	\resizebox{.91\hsize}{!}{$\begin{aligned}
		\psi\!=\, &2P_{1}^2P_{2}^2P_{3}^2\!-\!4P_{1}^2P_{2}P_{3}^2\!+\!2P_{1}^2P_{2}P_{3}\!+\!2P_{1}^2P_{3}^2\!
		-\!2P_{1}^2P_{3}\!\\
		&\qquad\qquad\qquad-\!4P_{1}P_{2}^2P_{3}^2\!+\!4P_{1}P_{2}P_{3}^2\!-\!2P_{1}P_{2}P_{3}\!+\!2P_{2}^2P_{3}^2,
		\end{aligned}$}
	\end{equation}
	\begin{equation}
	\resizebox{.99\hsize}{!}{$\begin{aligned}
		\omega\!=\!-(P_{1}^2P_{2}^2P_{3}^2\!-\!2P_{1}^2P_{2}P_{3}^2\!+\!P_{1}^2P_{3}^2
		\!-\!2P_{1}P_{2}^2P_{3}^2\!+\!2P_{1}P_{2}P_{3}^2\!+\!P_{2}^2P_{3}^2),
		\end{aligned}$}
	\end{equation}
\end{subequations}
are defined for the notation simplicity.

After a careful observation on the right hand side (RHS) of (29), we can deduce that the sign of $\frac{\partial \bar{\Delta}}{\partial p}$ is only determined by the numerator
\begin{equation}
\kappa(p) = \chi \!\cdot\! p^2+\psi \!\cdot\! p+\omega,
\end{equation}
since the denominator is always large than zero. To determine the monotonicity of the function $\bar\Delta$, we need to investigate the properties of the quadratic function $\kappa(p)$ on the feasible set of $p$ (i.e., [0, 1]). 

Firstly, after some algebra manipulations, we note that
\begin{subequations}
	\begin{equation}
	\omega=-P_{3}^2\Big[P_{1}^2(P_{2}-1)^2+P_{2}\big(P_{2}+2P_{1}(1-P_{2})\big)\Big],
	\end{equation}
	\begin{equation}
	\psi^2-4\chi \omega=4P_{3}^2P_{2}(P_{3}-P_{1})(1-P_{1})(P_{1}+P_{2}-P_{1}P_{2})^2.
	\end{equation}
\end{subequations}
As $0<P_{1}<P_{2}<1$ and $0<P_{1}<P_{3}<1$, we have that $\omega<0$ and $\psi^2-4\chi \omega>0$. This means that the curve of $\kappa(p)$ always intersects the Y-axis at the negative half of the Y-axis and it always has two roots on the X-axis, which can be given by
\begin{subequations}
	\begin{equation}
	x_{1}=\frac{-\psi -\sqrt{\psi ^2-4\chi  \omega}}{2\chi},
	\end{equation}
	\begin{equation}
	x_{2}=\frac{-\psi +\sqrt{\psi ^2-4\chi  \omega}}{2\chi},
	\end{equation}
\end{subequations}
where $x_{2}>x_{1}$. 

To further characterize the shapes of the function $\kappa(p)$, we now investigate $\chi$ and $\psi$ for the following four possible cases:


1) When $\chi>0$ and $\psi>0$, we always have $x_{1}<0$ and $x_{2}>0$. Since $p$ is in the range of $[0, 1]$, we have two sub-cases: (a) $x_{2}<1$ and (b) $x_{2}>1$. We draw the possible shapes of $\kappa(p)$ versus $p$ for the sub-case (a) and sub-case (b) in Fig. 4 (a) and Fig. 4 (b), respectively.

From Fig. 4 (a), we can see that in the sub-case (a), $\kappa(p)<0$ holds for $p\in[0, x_{2}]$, and $\kappa(p)>0$ holds for $p\in(x_{2}, 1]$. This means that the function $\bar\Delta$ is decreasing for $p\in[0, x_{2}]$ and is increasing for $p\in(x_{2}, 1]$. As a result, the minimum value of $\bar\Delta$ is achieved at $p=x_{2}$.

From Fig. 4 (b), we can see that in the sub-case (b), $\kappa(p)<0$ holds for $p\in[0, 1]$. Therefore, the function $\bar\Delta$ is always decreasing for $p\in[0, 1]$, which indicates that the value of $\bar\Delta$ is minimized at $p=1$.

2) When $\chi>0$ and $\psi<0$, we always have $x_{1}<0$ and $x_{2}>0$. Since $p$ is in the range of $[0,1]$, we also have two sub-cases: (a) $x_{2}<1$ and (b) $x_{2}>1$. The possible shapes of $\kappa(p)$ versus $p$ for the sub-case (a) and sub-case (b) are depicted as Fig. 5 (a) and Fig. 5 (b), respectively. 

Similar to case 1), we can prove that the minimum value of $\bar\Delta$ is achieved by setting $p=x_{2}$ and $p=1$ for the case $x_{2}<1$ and $x_{2}>1$, respectively.

3) When $\chi<0$ and $\psi<0$, we always have $x_{1}<x_{2}<0$. Therefore, there only exists one possible shape of $\kappa(p)$ versus $p$ in this case, which is depicted as Fig. 6. 

From Fig. 6, it is straightforward to see that $\kappa(0)<0$ holds for $p\in [0, 1]$. This means that the function $\bar\Delta$ is always decreasing for $p\in [0, 1]$ in this case. As a result, the minimum value of $\bar\Delta$ is achieved at $p=1$.

4) When $\chi<0$ and $\psi>0$, we always have $x_{2}>x_{1}>0$. Therefore, the possible shapes of $\kappa(p)$ versus $p$ can be depicted as Fig. 7. According to (30b), after some algebra manipulations, $\psi$ can be rewritten as 
\begin{equation}
\psi\!=-2P_{3}(P_{1}\!+\!P_{2}\!-\!P_{1}P_{2})(P_{1}\!-\!P_{1}P_{3}\!-\!P_{2}P_{3}\!+\!P_{1}P_{2}P_{3}).
\end{equation}
As $\psi>0$, $0<P_{1}<P_{2}<1$ and $0<P_{1}<P_{3}<1$, we thus have $P_{1}-P_{1}P_{3}-P_{2}P_{3}+P_{1}P_{2}P_{3}<0$, which can be rewritten as
\begin{equation}
P_{1}<\frac{P_{2}P_{3}}{1-P_{3}+P_{2}P_{3}}. 
\end{equation}


To further validate this case, we investigate the value of $-\chi$ under the constraint of (35). We rewrite $-\chi$ as a quadratic function of $P_{1}$, which can be expressed as 
\begin{equation}
	-\chi(P_{1})=\lambda P_{1}^2+\mu P_{1}+\xi,
\end{equation}
where
\begin{subequations}
   \begin{equation}
   	\lambda = P_{2}^2P_{3}^2-2P_{2}P_{3}^2+2P_{2}P_{3}-P_{2}+P_{3}^2-2P_{3}+1,
   \end{equation}
   \begin{equation}
   	\mu=2P_{2}P_{3}^2(1-P_{2})+P_{2}(1-P_{3}),
   	\end{equation}
   	\begin{equation}
   	\xi= P_{2}P_{3}(P_{2}P_{3}-1),
   	\end{equation}
\end{subequations}
are defined for the simplicity of notations.

As $0<P_{1}<P_{2}<1$ and $0<P_{1}<P_{3}<1$, according to (37b) and (37c), we have $\mu>0$ and $\xi<0$. In addition, if we set $P_{1}=1$, after some manipulations, we can attain that $-\chi({P_{1}})=(P_{3}-1)^2>0$. Similarly, if we substitute $P_{1}=\frac{P_{2}P_{3}}{1-P_{3}+P_{2}P_{3}}$ into the expression of $-\chi(P_{1})$, we can have $-\chi (P_{1})=\frac{P_{2}P_{3}(P_{2}-1)(P_{3}-1)^2}{(P_{2}P_{3}-P_{3}+1)^2}<0$. Based on the above analysis, we can draw the possible shapes of $-\chi(P_{1})$ versus $P_{1}$ for $\lambda>0$ and $\lambda<0$ in Fig. 8 (a) and Fig. 8 (b), respectively. Note that in these figures, we also show the possible positions of the point with X-coordinate being $\frac{P_{2}P_{3}}{1-P_{3}+P_{2}P_{3}}$.

\begin{figure}[htbp]
	\centering
	\subfigure[When $x_{2}<1$]{\includegraphics[height=2.8cm]{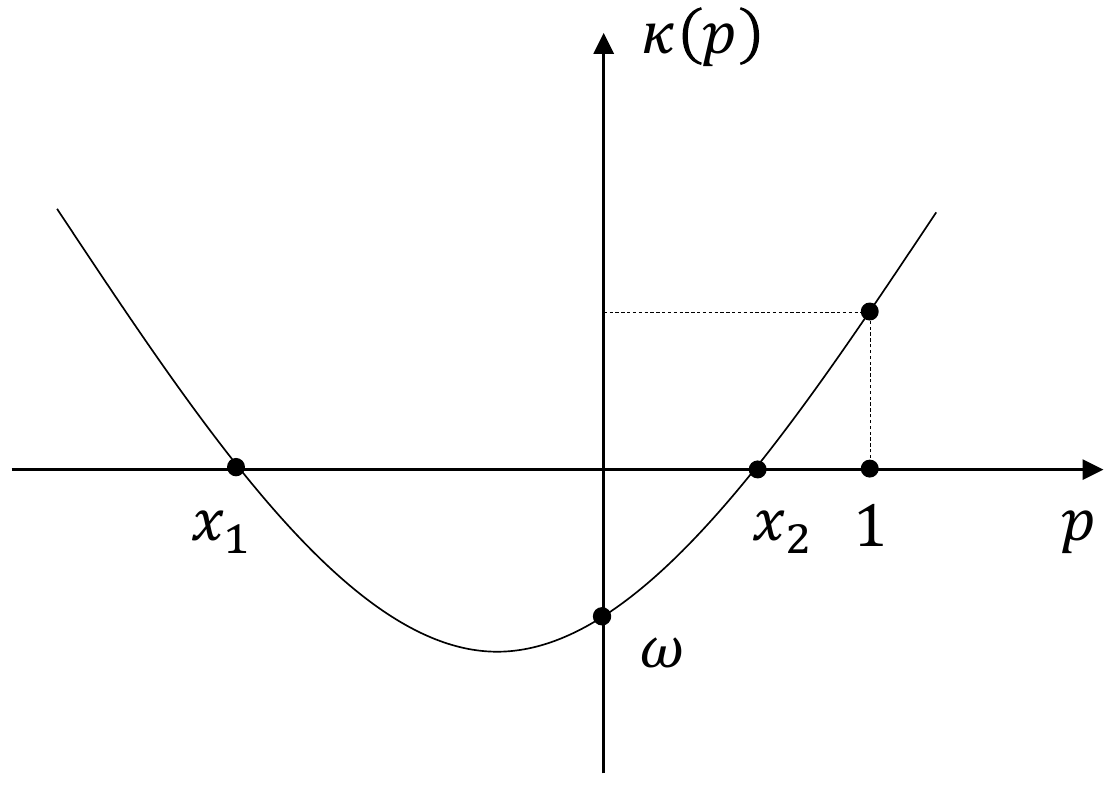}}
	\qquad
	\subfigure[When $x_{2}>1$]{\includegraphics[height=2.8cm]{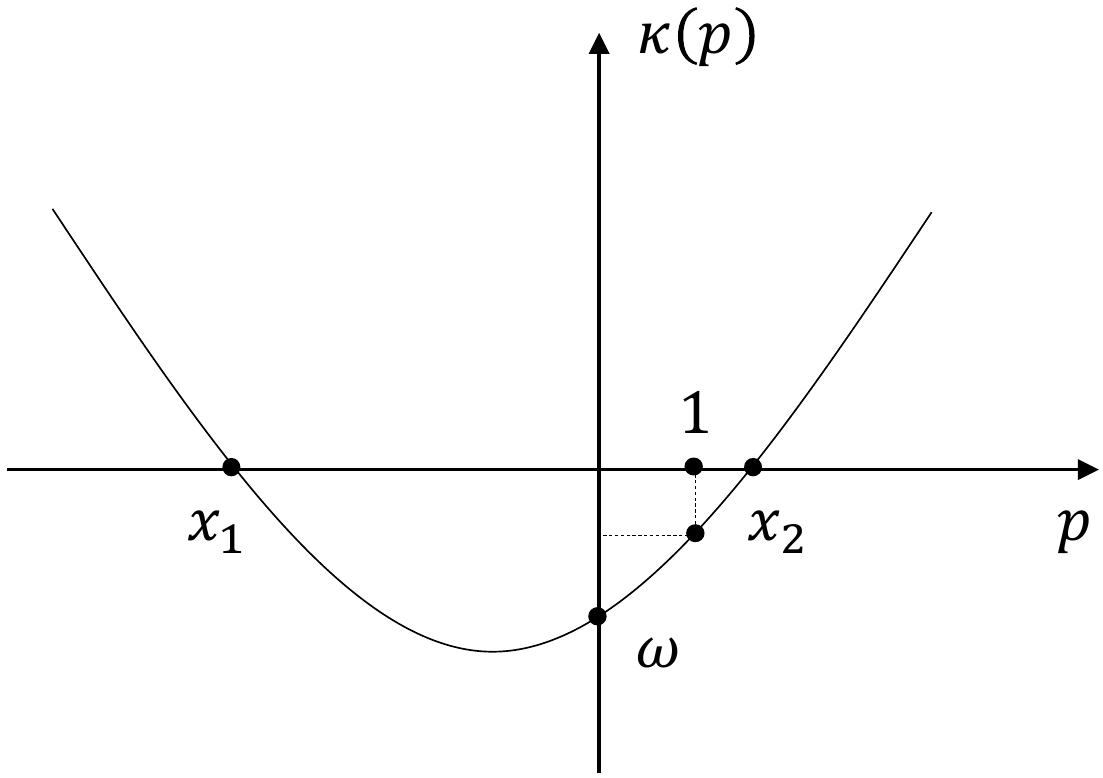}}
	\caption{Possible shapes for the function $\kappa(p)$ versus $p$ when $\chi>0$ and $\psi>0$.}
	\label{Fig. 4}
\end{figure}

\begin{figure}[htbp]
	\centering
	\subfigure[When $x_{2}<1$]{\includegraphics[height=2.8cm]{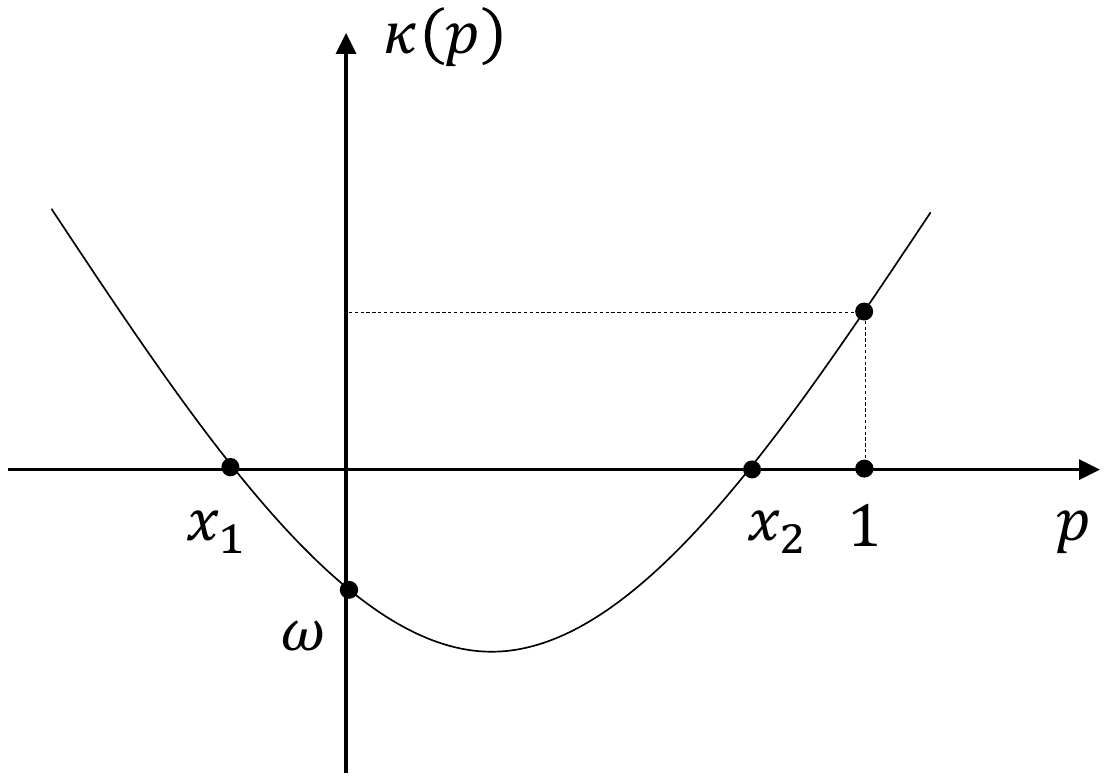}}
	\qquad
	\subfigure[When $x_{2}>1$]{\includegraphics[height=2.8cm]{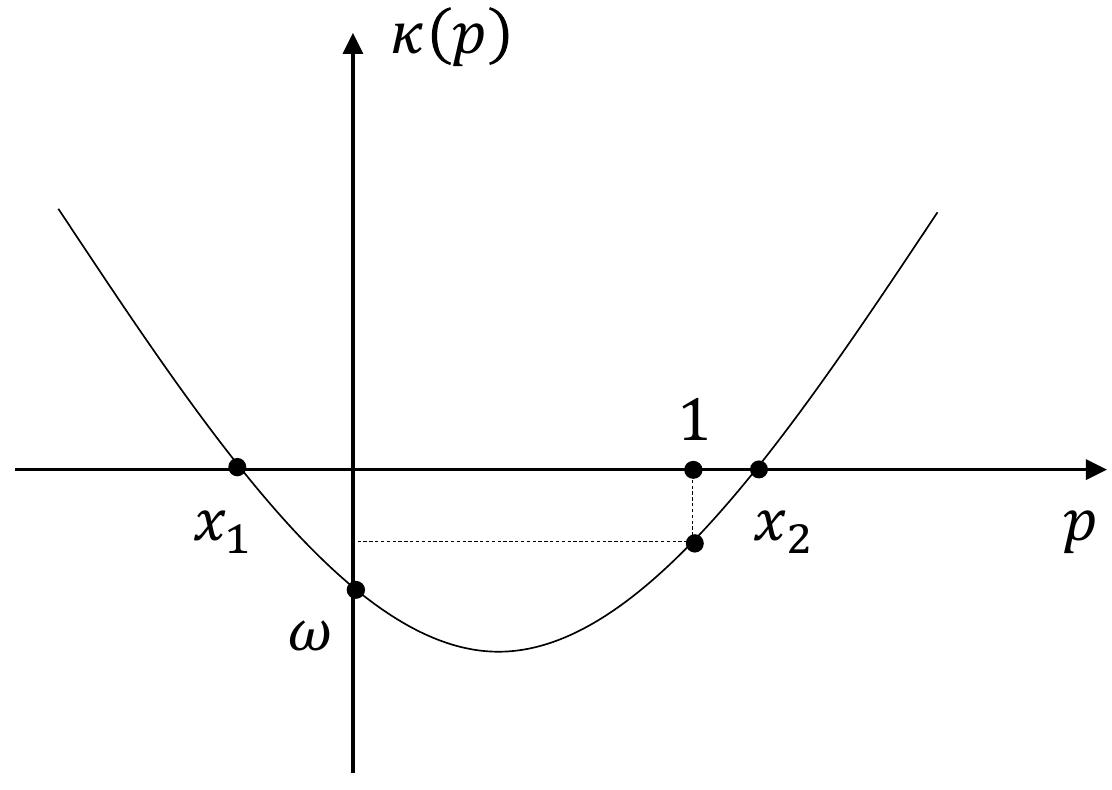}}
	\caption{Possible shapes for the function $\kappa(p)$ versus $p$ when $\chi>0$ and $\psi<0$.}
	\label{Fig. 4}
\end{figure}

\begin{figure}[tbp]
	\centerline{\includegraphics[height=5cm]{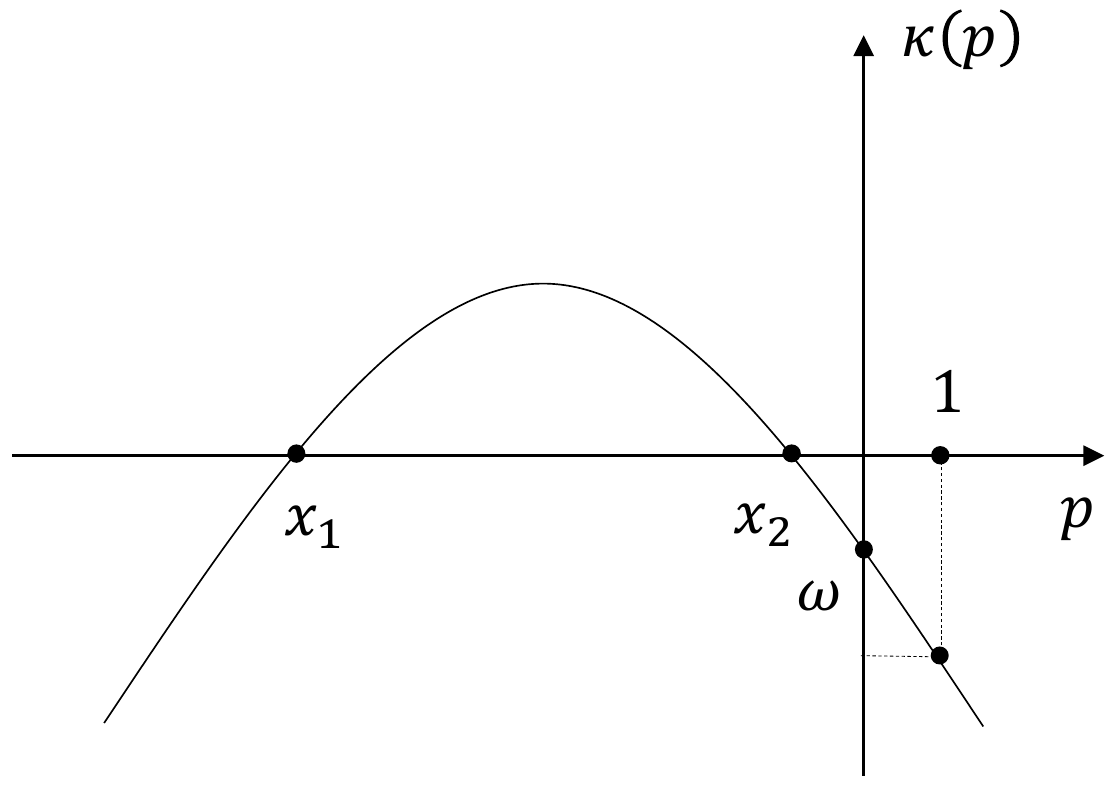}}
	\vspace{-0.7 em}
	\caption{Possible shapes for the function $\kappa(p)$ versus $p$ when $\chi<0$ and $\psi<0$.}
	\label{fig:two-hoop-relay-model-1}
\end{figure}

\begin{figure}[htbp]
	\centering
	\subfigure[When $x_{2}<1$]{\includegraphics[height=2.8cm]{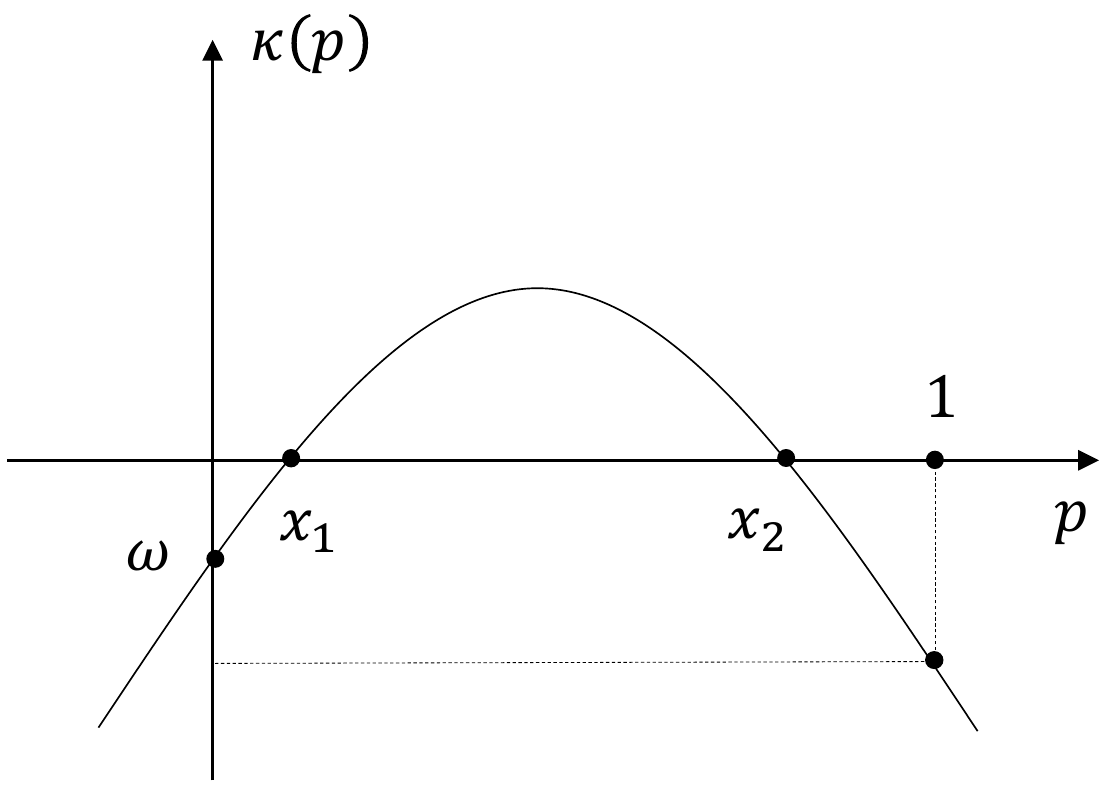}}
	\qquad	\subfigure[When $x_{2}>1$]{\includegraphics[height=2.8cm]{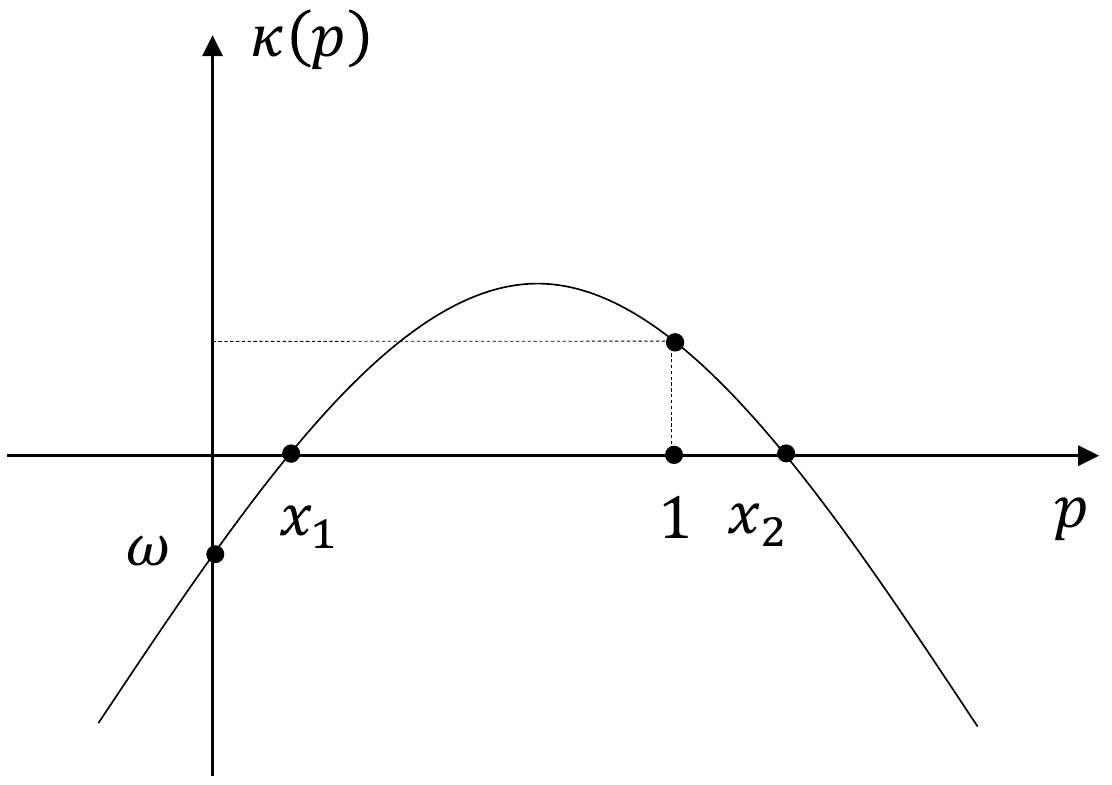}}
	\caption{Possible shapes for the function $\kappa(p)$ versus $p$ when $\chi<0$ and $\psi>0$.}
	\label{Fig. 4}
\end{figure}

\begin{figure}[htbp]
	\centering
	\subfigure[When $\lambda>0$]{\includegraphics[height=2.6cm]{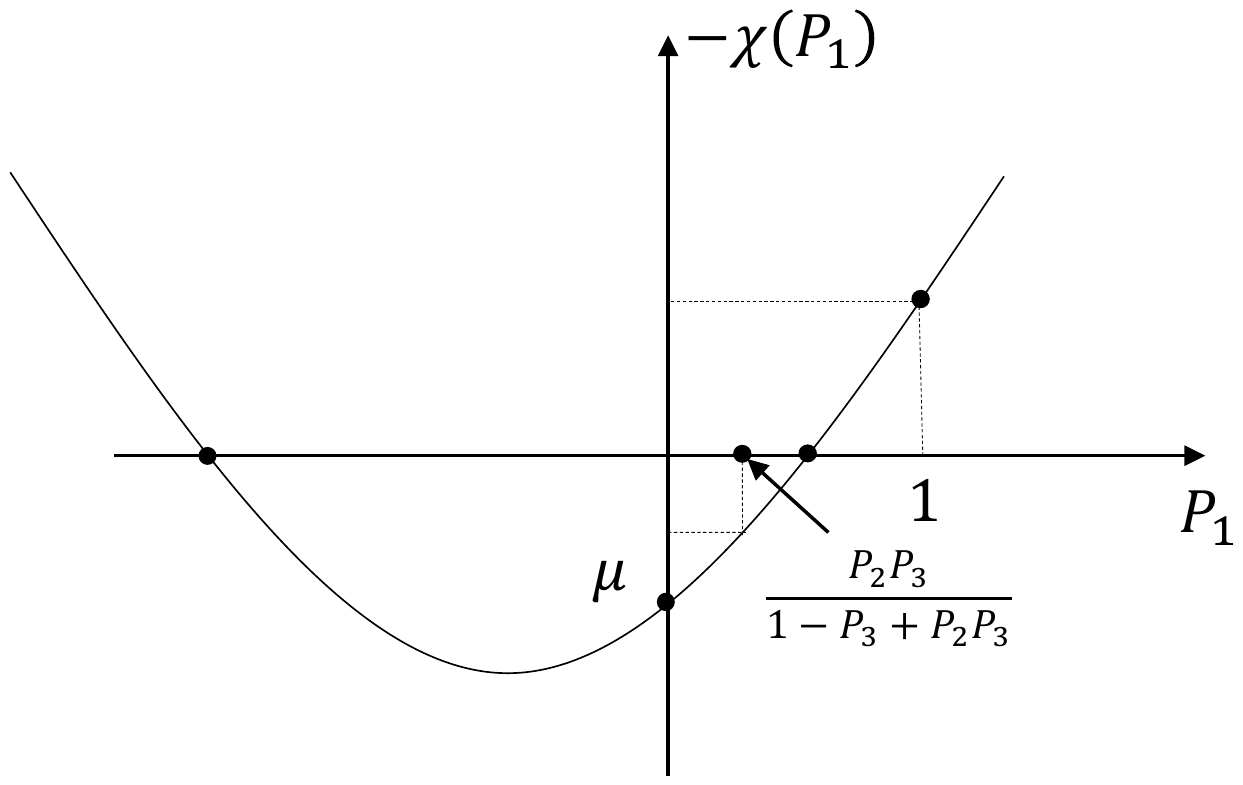}}
	\qquad
	\subfigure[When $\lambda<0$]{\includegraphics[height=2.6cm]{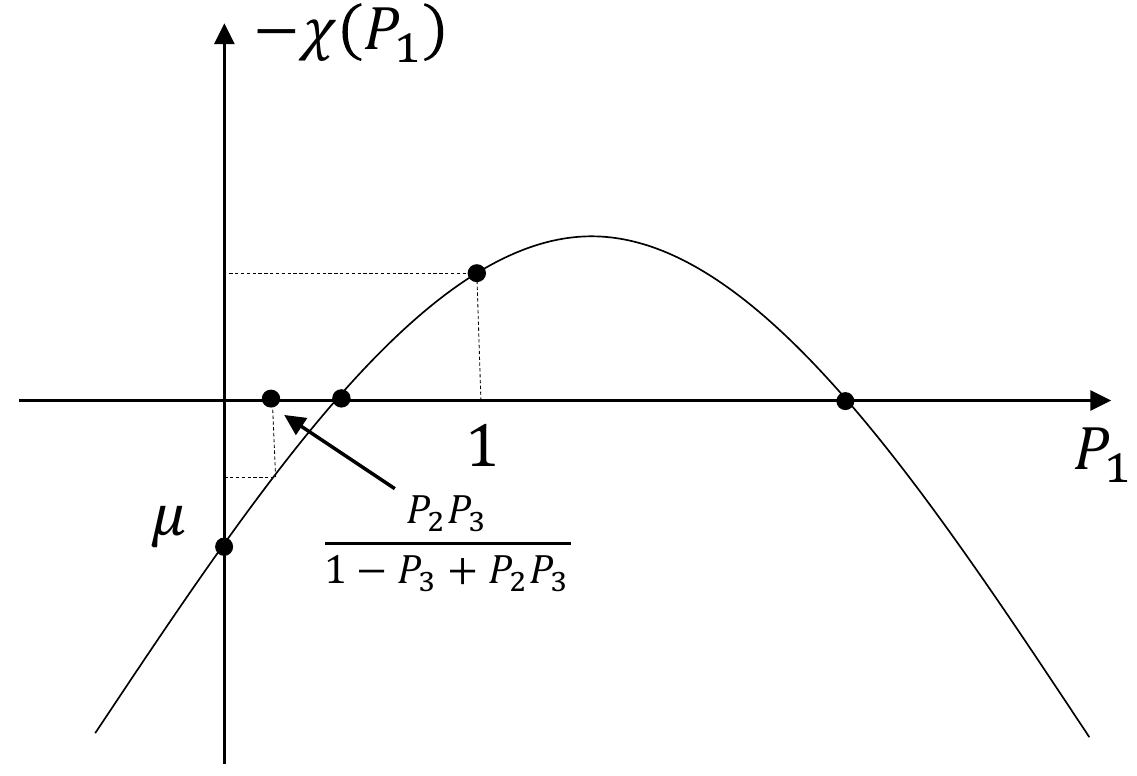}}
	\caption{Possible shapes for the function $-\chi(P_{1})$ versus $P_{1}$.}
	\label{Fig. 4}
\end{figure}

From Fig. 8 (a) and Fig. 8 (b), we can see that if $P_{1}<\frac{P_{2}P_{3}}{1-P_{3}+P_{2}P_{3}}$ and $0<P_{1}<1$,  $-\chi<0$ always holds. This means that we always have $\chi>0$ if $\psi>0$. This is contrary to the initial assumptions of this case (i.e., $\chi<0$ and $\psi>0$). Therefore, the case 4) is not valid.

In conclusion, we can find that in all of the possible case 1), case 2) and case 3), the minimum value of $\bar\Delta$ is achieved by setting $p=x_{2}$ and $p=1$ when $x_{2}<1$ and $x_{2}>1$, respectively. Therefore, based on (30a), (30b), (30c) and (33b), after some algebra manipulations, the optimal $p$ that minimizes $\bar\Delta$ can be given by (27).

This completes the proof of Theorem 1.

\end{document}